# Role of interstitial "*caged*" Fe in the superconductivity of FeTe$_{1/2}$Se$_{1/2}$

Anuj Kumar[1], Anand Pal[1], R. P. Tandon[2] and V. P. S. Awana[1*]

*Quantum Phenomena Application Division, National Physical Laboratory (CSIR)
Dr. K. S. Krishnan Road, New Delhi-110012*

*Department of Physics and Astrophysics, University of Delhi, New Delhi-110007*

All samples are synthesized through standard solid state reaction route and are quenched to room temperature systematically at 700$^0$C, 500$^0$C, 300$^0$C and room temperature (RT); named as 700Q, 500Q, 300Q and RTQ respectively. The structural and magnetic properties are studied. Careful Reitveld analysis of XRD patterns revealed that though all samples except 700Q are crystallized in single phase with space group *P4/nmm*, the presence of interstitial Fe (Fe$_{int}$) at *2c* site is increased from 5% for RTQ to 8% for 500Q. The 700Q sample is crystallized in Fe$_7$Se$_8$ phase. The transport and magnetization results revealed that though RTQ and 300Q are superconducting at 10 K and 13 K respectively, while the 500Q and 700Q are not. Magnetic ordering (T$_{mag}$) is observed at around 125 K for all the samples. The prominence of T$_{mag}$ in terms of effective moment is sufficiently higher for 500Q and 700Q than RTQ and 300Q. Summarily it is found that quenching induced disorder affects the occupancy of interstitial Fe in FeTe$_{1/2}$Se$_{1/2}$ and thus both its superconducting and magnetic properties. Further it clear that limited presence of interstitial Fe at *2c* site is not fully against observation of superconductivity, because 300Q sample possesses higher T$_c$ (13 K) for higher Fe$_{int}$ (6%) than RTQ sample with relatively lower T$_c$ (10 K) having lower Fe$_{int}$ (5%). Further the 500Q sample with much higher Fe$_{int}$ (8%) is though non-superconducting.

**INTRODUCTION**

The discovery oxy-pnictides with general structural formula REFeAsO$_{1-x}$F$_x$ (RE = rare earth) [1] of new iron-based high T$_c$ superconductors has ignited immense excitement in the material science community from past two years. The highest T$_c$ in these oxy-pnictides is reported at 56 K for SmFeAsO$_{0.85}$ and Gd$_{0.8}$Th$_{0.2}$FeAsO [2, 3]. Another important series of Iron based same family compounds are chalcogenides (FeSe, FeSe$_{1-\delta}$, FeSe$_{1-x}$Te$_x$, FeTe$_{1-x}$Se$_x$) having similar FeSe layers [4, 5]. The self doped anion deficient binary superconductor β-FeSe$_{1-\delta}$ shows a T$_c$ of 8 K which increases to 27 K by applying a hydrostatic pressure of 1.48 GPa [6]. This $dT_c/dP$ of around 9.1 K/GPa is the highest pressure effect value ever reported for any superconductor and it has attracted tremendous interest in the scientific community [7]. The $T_c$ of such superconductors barely decreases with applied field [8]. Besides external pressure [6, 7], the superconducting transition also increases by chemical pressure with Se at Te site substitution [8, 9]. The compound has similar structure and band filling as that of FeAs layer found in the quaternary iron arsenide and therefore presents a simple suitable model to study the interplay of structure, magnetism and superconductivity within iron based superconducting family [9]. There are some reports of decrease in superconducting transition ($T_c$) with substitution on Fe site by magnetic Ni and Co [10].

One of the important issues raised recently is the presence of interstitial Fe in the $FeSe_{1/2}Te_{1/2}$ unit cell [5, 11, 12]. Basically, the Fe does not occupy fully (100%) its designated *2a* site in *P4/nmm* structure, but also in relatively small quantity (5-8%) at *2c* site in stoichiometric compound without excess Fe compositions [13, 14]. It is believed that amount of interstitial Fe ($Fe_{int}$) affects the superconductivity and magnetism of the parent FeSe or FeSe/Te compounds [5, 11-14]. Interestingly in $FeTe_{1/2}Se_{1/2}$, though the substitution of Se is on site at Te, the coordinates of the Se and Te are slightly shifted [15].Generally, the interstitial Fe at 2c site is included in the stoichiometric formula as excess amount of Fe and its effect on physical properties are studied [5, 11,12]. We present here the control of interstitial Fe in $FeSe_{1/2}Te_{1/2}$ by quenching the nominal compound from various temperatures. We could vary the $Fe_{int}$ amount from 5% to 8% within same *P4/nmm* structure and studied its effect on superconductivity and magnetism.

**RESULTS AND DISCUSSION**

Samples of $FeTe_{1/2}Se_{1/2}$ are synthesized through standard solid state reaction route. The stoichiometric ratio of highly pure (> 3N) Fe, Se, and Te are ground, pelletized and then encapsulated in an evacuated ($10^{-3}$ Torr) quartz tube. The encapsulated tube is then heated at 750 $^o$C for 12 hours and slowly cooled to room temperature, this sample is named RTQ. Other samples were though synthesized at same temperature of 750 $^o$C, but the sealed quartz tube is quenched from 300 $^o$C, 500 $^o$C and 700 $^o$C to room temperature named as 300Q, 500Q and 700Q respectively. The X-ray diffraction (XRD) was performed at room temperature in the range of $10^o$-$60^o$ in equal 2θ step of $0.02^o$ using *Rigaku* diffractometer with *Cu $K_α$* (λ = 1.54Å) radiation. Riveted analysis was performed using the *FullProf* program. The AC and DC magnetization are performed on a physical property measurement system (PPMS-14T) from Quantum Design USA.

Figure 1 shows the XRD patterns of various temperatures (RT, 500$^o$C and 700$^o$C) quenched $FeTe_{1/2}Se_{1/2}$. Samples quenched at 500$^o$C and RT has similar XRD patterns, while the 700$^o$C quenched sample crystallizes in different phase ($Fe_7Se_8$) [16]. The 500$^o$C, 300$^o$C and RT samples are all crystallized in tetragonal structure with space group *P4/nmm*, without any detectable impurity within X-ray resolution limit. The diffraction pattern line arising from interstitial Fe at *2c* site appears next to main diffraction peak and is marked with #. It is clear that the intensity of $Fe_{int}$ arising peak (#) is increased for 500Q sample. As far as the details of fitting parameters are concerned, not only the Se and Te coordinate (z) is allowed to vary as per ref. 15, the Fe(2a) and Fe(2c) occupancies are also left free. Interestingly, though the sum of Fe(2a) and Fe(2c) sitting Fe fits close to the nominal (1.0), the fact is that we left the occupancies of the two floating and not fixed during the analysis. The best fits within $\chi^2$ range of less than 2 are obtained. Though clearly the $Fe_{int}$ occupancy increase from 5% to 8%, the z coordinates shifts of Se/Te is not remarkable from one sample to another.

Detailed Reitveld parameters including coordinate positions, lattice parameters and different atom occupancies in particular of $Fe_{int}$ for RTQ, 300Q, and 500Q samples are given in Table 1-3. The lattice parameters for the RTQ sample are in good agreement with earlier reports [7-9, 11-15]. In comparison to RTQ sample, the *c* lattice parameter first increases for 300Q sample and later decreases for 500Q sample. Further, the $Fe_{int}$ occupancy increase from 5% to 8% for RTQ to 500°C samples respectively. This in accordance with increasing intensity of the # marked diffraction line in Figure 1. As mentioned in the introduction itself, the $Fe_{int}$ is the part of unit cell itself, but occupying *2c* site instead of its usual *2a* site. It is concluded from the XRD results that the RTQ, 300Q and 500Q samples are crystallized in usual *P4/nmm* space group while the 700Q sample has changed its phase. It is clear that amount of $Fe_{int}$ increases for higher temperature quenched samples than RTQ.

The AC magnetic susceptibility of RTQ and 300Q $FeTe_{1/2}Se_{1/2}$ samples are given in Figure 2. The RTQ and 300Q samples exhibit superconducting transition at 10 K and 13 K respectively. The superconducting transitions are seen in imaginary part of AC magnetic susceptibility as well, being higher for 300Q sample than RTQ. Interestingly, as discussed in previous section detailing XRD results, the amount of $Fe_{int}$ is slightly more (6%) in 300Q sample than RTQ (5%). On the other hand the $T_c$ is higher (13 K) for 300Q sample than 10 K for the RTQ. It seems the increasing amount of $Fe_{int}$ is not detrimental to the superconductivity of $FeSe_{1/2}Te_{1/2}$. However this will be clear after we discuss the superconductivity of 500Q sample, which possess even higher (8%) $Fe_{int}$.

The DC magnetic susceptibility of RTQ and 300Q samples at 20 Oe in both field cooled (FC) and zero field cooled (ZFC) situations is depicted in Figure 3. Though the bulk superconducting transition temperature ($T_c$) is seen in both FC and ZFC at same temperature as being observed in AC susceptibility (Figure 2), the normal state DC moment is slightly different. It is clear from Figures 2 and 3 that the 300Q sample exhibits higher $T_c$ in comparison to RTQ. The DC magnetic susceptibility of high temperatures (500Q and 700Q) samples are shown in inset of Figure 4. Both the 500Q and 700Q samples are not superconducting. Further clear magnetic transitions (clear FC and ZFC separation) are seen at 125 K and 225 K respectively for the 500Q and 700Q samples respectively. In fact for 700Q sample the opening of FC and ZFC is right up to 300 K. This is because the magnetic transition temperature for $Fe_7Se_8$ is reported above room temperature (460 K) [16]. As far as the 125 K major transition is concerned for both 500Q and 700Q samples, the same is intriguing because it matches with the Verway transition temperature (120 K) of Fe ions in $Fe_3O_4$ [17,18]. At this point, it is difficult for us to assign any $Fe_3O_4$ impurity in our 500Q sample within the XRD detection limit. As seen in Figure 1, both the RTQ and 500Q samples possess identical XRD patterns within same space group.

The normal state moments of both the 500Q and 700Q samples are larger than the superconducting 300Q and RTQ samples. There is possibility that minute amount of $Fe_3O_4$ has contaminated both 500Q and 700Q samples, however within XRD resolution we discard this possibility at present. Certainly more specific studies with much better resolution such as synchrotron and high resolution electron microscopy are required to resolve this issue. With an optimistic approach based on XRD results it seems that that increased amount of $Fe_{int}$ contributes to the total normal state magnetic susceptibility of the $FeSe_{1/2}Te_{1/2}$ system. Though the 700Q sample results cannot be discussed in direct comparison to RTQ, 300Q, and 500Q samples because it crystallizes in a different crystallographic phase, the other three can be compared directly. It is clear that with increase in quenching temperature from RTQ to 300Q and 500Q the $Fe_{int}$ increases from 5% to 6% and 8%, on the other hand the superconductivity first increases from 10 K to 13 K and later to non superconductivity. Also the higher temperature (500Q) quenched sample exhibits magnetic ordering and higher Fe moment without superconductivity. This is evidenced from the isothermal magnetization (*MH*) loops for optimum superconducting (13 K) 300Q and non-superconducting (500Q) samples, being shown in Figures 4(a) and (b). 300Q sample shows typical superconducting loop at 2 K riding over possible magnetic background of $Fe_{int}$. Above superconducting transition temperature (50 K, 100 K, 150 K and 200 K), the hysteresis loops are clearly of ferromagnetic nature, with saturation moment of around $0.02\mu_B$/Fe. In fact the normal state (above $T_c$) isothermal magnetization plots are only scant in literature to judge if $Fe_{int}$ contributes to the net magnetization of this class of compounds. It is warranted to know the exact magnetic nature of $Fe_{int}$ in FeSe/Te superconductor. In current article we have raised an important issue related to control of $Fe_{int}$ without invoking nominal excess Fe and the effect of same is studied the superconducting properties of stoichiometric $FeSe_{1/2}Te_{1/2}$ compound.

The authors from NPL would like to thank Prof. R.C. Budhani (Director, NPL) for his keen interest in the present work. One of us, Anuj Kumar, would also thank Council of Scientific and Industrial Research (*CSIR*), New Delhi for financial support through Senior Research Fellowship (*SRF*).

**Table 1 FeTe$_{1/2}$Se$_{1/2}$ RTQ: Lattice Parameters and Unit cell volume:**
$a$ = 3.8009(7) Å, $c$ = 6.0287(7) Å, Vol. 87.100 Å$^3$ and $\chi^2$ = 1.49

| Atom | Site | $x$ | $y$ | $z$ | Occupancy |
|---|---|---|---|---|---|
| Fe | 2a | 0.7500 | 0.2500 | 0.0000 | 0.951 |
| Se | 2c | 0.2500 | 0.2500 | 0.2425(2) | 0.500 |
| Te | 2c | 0.2500 | 0.2500 | 0.2894(1) | 0.500 |
| Fe$_{int}$ | 2c | 0.2500 | 0.2500 | 0.7824(3) | 0.051 |

**Table 2 FeTe$_{1/2}$Se$_{1/2}$ 300Q: Lattice Parameters and Unit cell volume:**
$a$ = 3.7945(2) Å, $c$ = 6.0311(4) Å, Vol. 86.839 Å$^3$ and $\chi^2$ = 2.09

| Atom | Site | $x$ | $y$ | $z$ | Occupancy |
|---|---|---|---|---|---|
| Fe | 2a | 0.7500 | 0.2500 | 0.0000 | 0.944 |
| Se | 2c | 0.2500 | 0.2500 | 0.2609(9) | 0.500 |
| Te | 2c | 0.2500 | 0.2500 | 0.2775(2) | 0.500 |
| Fe$_{int}$ | 2c | 0.2500 | 0.2500 | 0.8102(8) | 0.063 |

**Table 3 FeTe$_{1/2}$Se$_{1/2}$ 500Q: Lattice Parameters and Unit cell volume:**
$a$ = 3.7959(1) Å, $c$ = 6.0092(4) Å, Vol. 86.587 Å$^3$ and $\chi^2$ = 1.78

| Atom | Site | $x$ | $y$ | $z$ | Occupancy |
|---|---|---|---|---|---|
| Fe | 2a | 0.7500 | 0.2500 | 0.0000 | 0.923 |
| Se | 2c | 0.2500 | 0.2500 | 0.2405(5) | 0.500 |
| Te | 2c | 0.2500 | 0.2500 | 0.2937(9) | 0.500 |
| Fe$_{int}$ | 2c | 0.2500 | 0.2500 | 0.8179(1) | 0.082 |

# FIGURE CAPTIONS

**Figure 1** XRD patterns of RTQ, 500Q and 700Q FeSe$_{1/2}$Te$_{1/2}$ samples at room temperature

**Figure 2** AC magnetic susceptibility versus temperature plots for superconducting RTQ and 300Q FeSe$_{1/2}$Te$_{1/2}$ samples

**Figure 3** DC magnetic susceptibility versus temperatures plots for 300Q and RTQ FeSe$_{1/2}$Te$_{1/2}$ samples at 20 Oe field; the inset shows the same for 500Q and 700Q samples.

**Figure 4** Typical magnetization loops at different temperatures for (a) 300Q and (b) 500Q samples as function of applied magnetic field in the range of - 20 kOe to + 20 kOe

Figure 1

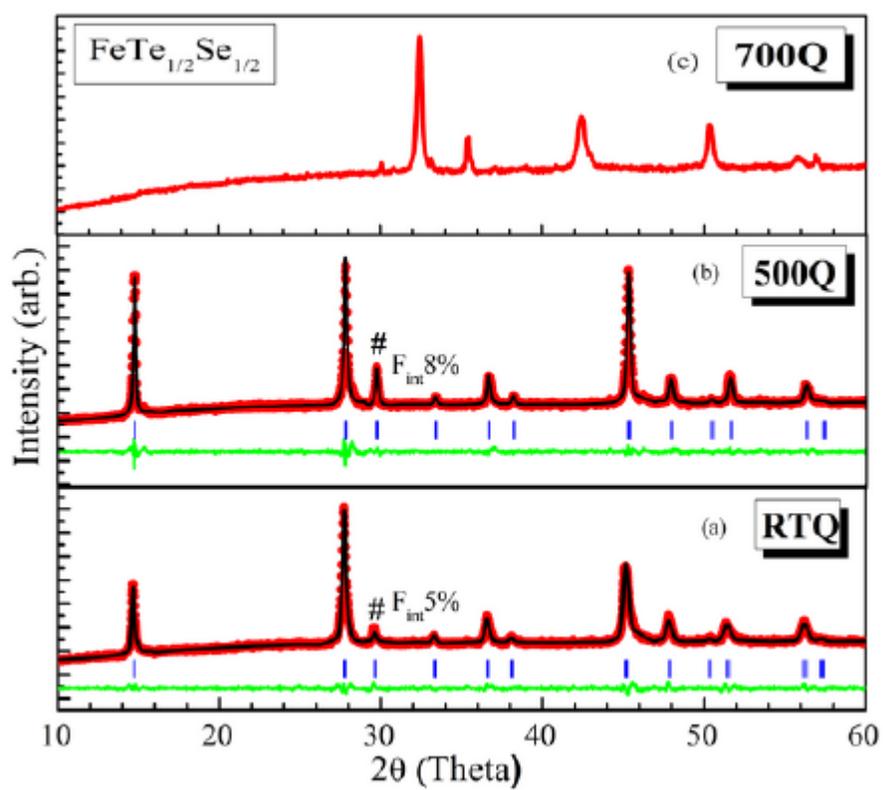

Figure 2

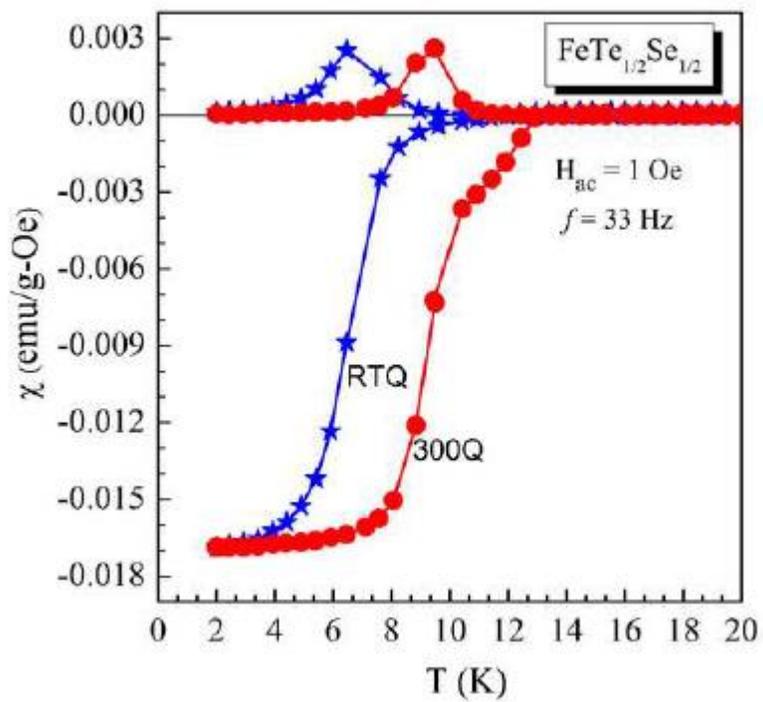

Figure 3

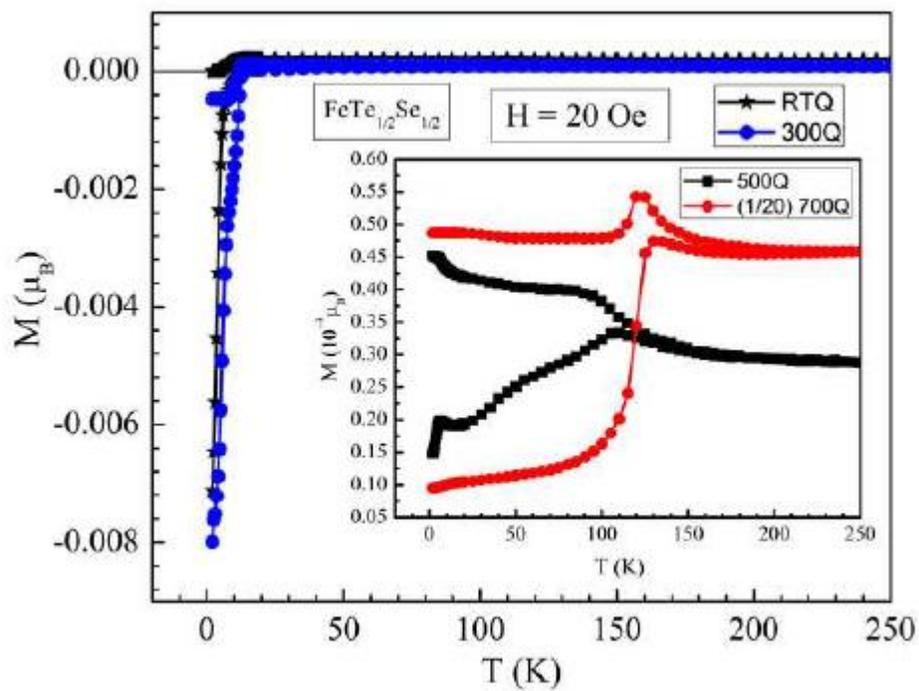

Figure 4(a) and 4(b)

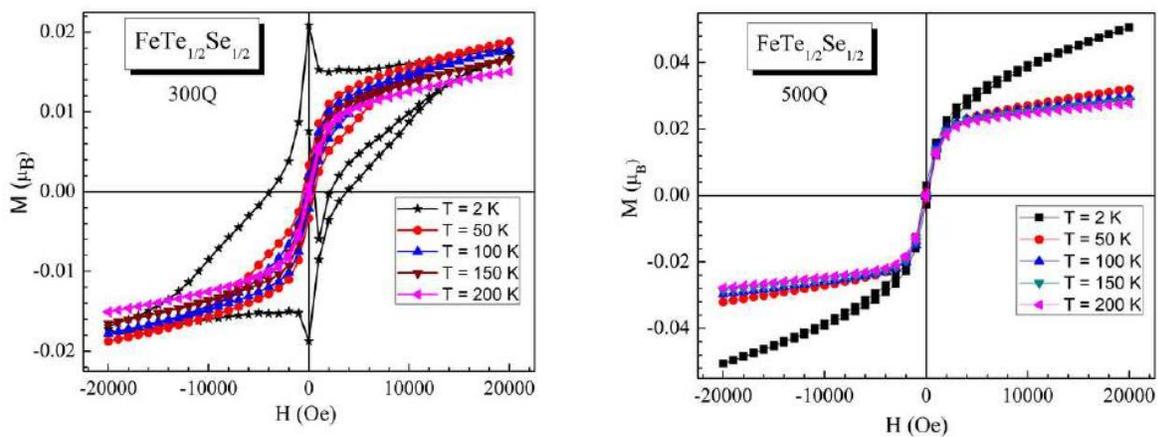

8